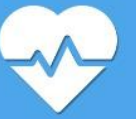

# Anemia, weight, and height among children under five in Peru (2007-2022): a panel data analysis

# Anemia, peso e altura entre crianças menores de cinco anos no Peru (2007-2022): uma análise de dados de painel

# Anemia, peso y talla en los niños hasta cinco años de edad en Perú (2007-2022): un análisis con datos de panel

**Luis-Felipe Arizmendi**
PhD in Economics
Institution: Faculty of Health Sciences, Universidad Alfonso X El Sabio
Address: Madrid, Spain
E-mail: larizech@uax.es

**Carlos de la Torre Domingo**
PhD in Health Sciences, Physiotherapy
Institution: Faculty of Health Sciences, Universidad Alfonso X El Sabio
Address: Madrid, Spain
E-mail: carlost@uax.es

**Erick W. Rengifo**
PhD in Economics
Institution: Deparment of Economics and Center for International Policy Studies (CIPS), Fordham University
Address: Bronx, New York, United States of America
E-mail: rengifomina@fordham.edu

## ABSTRACT

Panel data methods are becoming crucial in econometrics and social sciences for analyzing data that follows the same entities (such as individuals, firms, countries) over multiple time periods. In this article, we employ a useful approach (Feasible Generalized Least Squares) to assess if there are statistically relevant relationships between hemoglobin (adjusted to sea-level), weight, and height for the period 2007–2022 in children up to five years of age in Peru. By using this method, we may find a tool that allow us to confirm if the relationships considered between the target variables by the Peruvian agencies and authorities are in the right direction to fight against chronic malnutrition and stunting. While health policies aimed at reducing anemia and improving nutrition, such as supplementation programs and maternal education initiatives, may have contributed to these trends, further research with multidisciplinary collaboration is necessary to quantify the specific impact of these interventions on the observed outcomes.

**Keywords:** Anemia. Children. Anthropometrics. Hemoglobin. Panel Data.

**RESUMO**
Os métodos de dados de painel estão se tornando cruciais na econometria e nas ciências sociais para analisar dados que acompanham as mesmas entidades (como indivíduos, empresas, países) em vários períodos de tempo. Neste artigo, empregamos uma abordagem útil (Feasible Generalized Least Squares) para avaliar se há relações estatisticamente relevantes entre hemoglobina (ajustada ao nível do mar), peso e altura no período de 2007 a 2022 em crianças de até cinco anos de idade no Peru. Ao usar esse método, podemos encontrar uma ferramenta que nos permita confirmar se as relações consideradas entre as variáveis-alvo pelos órgãos e autoridades peruanos estão na direção certa para combater a desnutrição crônica e o atraso no crescimento. Embora as políticas de saúde destinadas a reduzir a anemia e melhorar a nutrição, como programas de suplementação e iniciativas de educação materna, possam ter contribuído para essas tendências, são necessárias mais pesquisas com colaboração multidisciplinar para quantificar o impacto específico dessas intervenções nos resultados observados.

**Palavras-chave:** Anemia. Crianças. Antropometria. Hemoglobina. Dados de Painel.

**RESUMEN**
Los métodos de datos de panel se están volviendo cruciales en la econometría y las ciencias sociales para analizar datos que siguen a las mismas entidades (como individuos, empresas, países) durante múltiples períodos de tiempo. En este artículo, empleamos un enfoque útil (mínimos cuadrados generalizados factibles) para evaluar si existen relaciones estadísticamente relevantes entre la hemoglobina (ajustada al nivel del mar), el peso y la altura para el período 2007-2022 en niños de hasta cinco años de edad en Perú. Mediante el uso de este método, podemos encontrar una herramienta que nos permita confirmar si las relaciones consideradas entre las variables objetivo por los organismos y autoridades peruanos van en la dirección correcta para luchar contra la desnutrición crónica y el retraso en el crecimiento. Si bien las políticas sanitarias destinadas a reducir la anemia y mejorar la nutrición, como los programas de suplementación y las iniciativas de educación materna, pueden haber contribuido a estas tendencias, es necesario seguir investigando con una colaboración multidisciplinar para cuantificar el impacto específico de estas intervenciones en los resultados observados.

**Palabras clave:** Anemia. Niños. Antropometría. Hemoglobina. Datos de Panel.

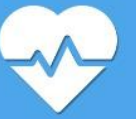

# 1 INTRODUCTION

The primary objective of this study is to analyze trends in hemoglobin levels, weight, and height of children under five in Peru over the period 2007–2022. By employing Feasible Generalized Least Squares (FGLS) panel data methods, we aim to assess whether the observed improvements align with expected developmental trends and explore the potential role of public health interventions in these trends. By doing this, we expect to cover throughout the article some important aspects of public health with recent quantitative techniques, exploring both theoretical and practical issues.

During the last twenty-five years, the Peruvian economy has experienced a quite prosperous business cycle, with substantial GDP growth in both senses, total and per capita, in real terms. As such, Peru is now classified as a developing country with upper middle level of income per capita by the United Nations (UN). Poverty levels, as related to population percentage, have dropped from almost 59% in 2004 to 20% in 2019 (prior to the COVID-19 global event). There is, however, a substantial part of its people still living in poverty. To reduce the consequences of such situation, particularly among children of the most vulnerable social classes and groups, the efforts of the Peruvian Government through various budgetary programs to reduce anemia levels in children under five years old have been noteworthy, especially to what seems to be accomplished between years 2000 to 2022. These programs include treatments or direct interventions based on nutrients administered by the mothers of these children or through public health centers.

There is, however, a persistent problem with chronic malnutrition and stunting in Peru. This last condition refers to the impairment of growth and development in children caused not only by chronic malnutrition, but recurrent infections, and insufficient psychosocial stimulation as well. When we combine studies of hemoglobin levels, and weight and height measurements, we may have the core indicators of inadequate nutrition during critical stages of growth (particularly within the first 1000 days of life), which span from conception to two years of age. This term is used to characterize children who have a significantly shorter height in relation to their age compared to a standard reference of a healthy population, implying chronic malnutrition. The issue of anemia and its connection to nutrition in children is closely related to stunting, an important indicator of malnutrition. Malnutrition not only hinders height growth but also has the potential to affect cognitive development and increase susceptibility to diseases such as anemia. Anemia is defined as a condition characterized by hemoglobin levels below WHO-defined thresholds, leading to reduced oxygen transport capacity in the blood. It is also characterized by a deficiency of healthy red blood cells necessary for efficient oxygen, that often arises from nutritional deficiencies, such as insufficient intake of iron and vitamin B12, among other essential nutrients. Consequently, addressing both stunting and anemia simultaneously is crucial in efforts to improve the health and nutrition of children. By tackling these challenges, we can ensure that children reach their maximum potential in terms of growth and development.

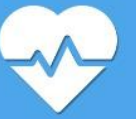

In the last two decades, there have been many research pieces related to the subject, trying to describe situations and conditions of malnourished children. Authors and practitioners (Amugsi *et al.*, 2020) in sub-Saharan countries, (Antenen and Van Geertruyden, 2021) for Ethiopia, (Dyness *et al.*, 2018) in Tanzania, (Ewusie *et al.*, 2014) for Ghana, (Al_Kassab-Córdova *et al.*, 2022) in Peru for rural and urban disparities, (Rahman *et al.*, 2019) for Bangladesh, and (Sunguya *et al.*, 2020) for Tanzania, deserve to be mentioned as the scholars that have provide us with the scope for the present work and to search for a different, hopefully original, way to search a deep understanding on the chronic anemia and how to overcome this problem when related to basic anthropometric indicators.

In the case of Peru, according to Alcazar (2013), it is an iron-deficient diet what the prevalent nutritional issue, as it is also recurrent in many developing countries. By recognizing the significance of this problem, the government has made it a priority to implement interventions aimed at reducing iron deficiency anemia. In relation to Peru, for instance, the objective is to enhance the nutritional status of our children and foster their future development. To address this issue, increased financial resources have been allocated through the Coordinated Nutritional Program (PAN) within the Budget for Results. Additionally, anemia prevention and treatment have been integrated into the Essential Plan for Health Insurance.

Curi-Quinto *et al.* (2019) conclude that malnutrition is a prevalent problem in Peru. They point out that malnutrition among children under five and women of reproductive age, shows a heterogenous distribution when compared to socio-economic indicators. As such, these authors support the implementation of a set of policies and interventions to overcome malnutrition, and to reduce disparities as well.

Finally, since near 40% of the Peruvian population lives in the Andean mountains, Gonzales *et al.* (2021) when addressing the cutoff hemoglobin (Hb) value of 11 g/dL for preschool-aged children set by the WHO and its adjusted-by-altitude measurements, conclude that adjusted hemoglobin values should be considered before 1000 m, paying attention to ranges adjusted to smaller groups for children and not the same reference range for children from 6 to 59 months. All this in mind, we emphasize the main objective of our research, that is, to analyze what has been happening on anemia and relevant anthropometric indicators of Peruvian children under five years of age for the period under study.

## 2 MATERIALS AND METHODOLOGY

The main source of information and data comes from the individuals and family members that have participated in the Demographic and Family Health Survey – ENDES (2000, 2022) the *Encuesta Demográfica y de Salud Familiar* (in Spanish), made annually by the National Institute of Statistics and Informatics – INEI or *Instituto Nacional de Estadística e Informática*, which is the government agency responsible for collecting, processing, analyzing, and disseminating statistical information in Peru.

Below is a concise summary of the methodology used by the INEI for the preparation, implementation, and analysis of the ENDES:

a) The questionnaire follows international standards and guidelines, particularly those provided by organizations like the United Nations and the World Health Organization (WHO). The questionnaire is tailored to address national health and demographic priorities, with sections covering fertility, child and maternal health, nutrition, vaccination, family planning, and domestic violence, among others. A pilot test is conducted to validate the content, structure, and language, ensuring cultural relevance and clarity;

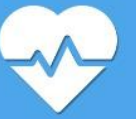

b) The survey employs a multistage, probabilistic, and stratified sampling method to ensure national and regional representativeness. The sampling frame is based on updated census data, with stratification by urban and rural areas, as well as by geographic regions. Sampling units include households and individuals, ensuring the selection of a statistically representative population;

c) Data collection is carried out through face-to-face interviews using trained enumerators equipped with digital devices to ensure accuracy and consistency. Enumerators follow a structured protocol, and responses are recorded directly in an electronic data collection system to minimize errors. Supervisory mechanisms are in place to ensure adherence to the methodology and maintain high data quality;

d) Enumerators and supervisors undergo rigorous training to familiarize themselves with the questionnaire, interview techniques, and ethical considerations, including the protection of respondents' confidentiality. Standardized manuals and simulation exercises are employed to ensure uniformity in data collection across regions;

e) The data collected is transferred to a centralized system for processing. The INEI employs statistical software to perform data cleaning, coding, and validation. Consistency checks and imputation techniques are applied to handle missing or inconsistent data;

f) Statistical analyses are conducted to produce estimates and indicators aligned with national and international frameworks. Results are disaggregated by age, gender, geographic location, and other relevant variables to support targeted policymaking. The findings are compiled into detailed reports and summary publications, which are disseminated to government agencies, researchers, and the public;

g) The INEI adheres to strict ethical guidelines to protect the rights and privacy of respondents. Informed consent is obtained from all participants, and data confidentiality is rigorously maintained throughout the survey process.

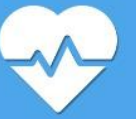

ENDES is designed to gather comprehensive data on demographics, health, and family life within the country. It plays a crucial role in informing public policy, health initiatives, and social programs. This includes information on fertility, mortality, family planning, maternal and child health, nutrition, and other health-related issues, especially those aimed at improving the health and welfare of the population. It was executed for the first time in 1996, although without measurements of hemoglobin and anemia, that were made in 2000. ENDES, as such, was reinstated in 2005, but that year and 2006 did not include tests for hemoglobin (no anemia evaluation). Since 2007 and on, ENDES always comprises blood testing, with a clear methodology (see 2018).

Although the number of participants can vary widely, ENDES aims to survey tens of thousands of households each year to ensure a robust dataset. The survey collects hundreds of thousands of observations across its various modules and questionnaires. The key variables and areas covered by ENDES, include: 1) Demographics: Age, sex, marital status, education, employment, and migration; 2) Health: Immunization rates, prevalence of diseases, access to healthcare services, and health behaviors; 3) Nutrition: Nutritional status of children and adults, breastfeeding practices, and food security; 4) Reproductive Health: Fertility rates, contraceptive use, antenatal and postnatal care, and birth history; 5) Child Health: Growth monitoring, development indicators, and morbidity and mortality among children under five; 6) Social and Economic Indicators: Household composition, housing conditions, access to basic services (cell phone penetration, for instance), and poverty indicators. The ENDES data used in this study is publicly available and adheres to strict anonymization protocols implemented by the National Institute of Statistics and Informatics (INEI) to protect participant confidentiality. This study did not involve human intervention or access to identifiable data, and thus, ethical committee approval was not required. All analyses comply with international ethical standards for research using secondary data.

Results are published in detailed reports, which include statistical tables, graphs, and thematic analysis. These open data reports are made available to the public, researchers, policymakers, and international organizations. ENDES data is a critical input for developing and adjusting health policies and programs in Peru. Data from ENDES also contributes to global databases, such as those maintained by the World Health Organization -WHO and the United Nations, facilitating international comparisons and global health monitoring. As such, ENDES is a cornerstone in Peru's statistical and health information system, providing invaluable insights into the well-being of its population and informing efforts to improve public health and social outcomes.

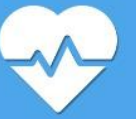

In this research, we use the variables Age in Months (HW1 as it is labelled in ENDES), Weight in kilograms (HW2), Height in centimeters (HW3), Level of Hemoglobin in g/dL, adjusted by sea-level (HW56), and the level of anemia (HW57) from the source ENDES. All missing values and outliers have been removed in this work. Outliers were identified based on domain-specific knowledge and logical thresholds. Extreme values deemed absurd or impossible for human conditions (e.g., unrealistic ages, biologically improbable weight or height measurements, or implausible health indicators) are flagged and removed. Statistical techniques, such as interquartile range (IQR) rules or z-scores, were used to detect extreme deviations, but our final decision was guided by context-specific criteria and expert judgment. In the case of the variable Hemoglobin, the formula for adjustment is the one developed by USAID-INACG (2002). All data is open and available online through the INEI-ENDES website [15] and it can be also provided by the authors upon request. Anemia severity is classified according to WHO standards: mild (Hb 10.0–10.9 g/dL), moderate (Hb 7.0–9.9 g/dL), and severe (Hb <7.0 g/dL). Chi-square tests confirm statistically significant declines in anemia prevalence across the years analyzed. Earlier variations in sample sizes reflect the evolving scope and scale of the ENDES surveys, with improved participation in later years enhancing data quality.

Feasible Generalized Least Squares (FGLS) is a statistical method used in econometrics to address issues of heteroscedasticity and serial correlation within panel data. Unlike Ordinary Least Squares (OLS), which assumes constant variance across observations, FGLS estimates different variances for different entities or over time, adapting to the variability observed in the dataset. FGLS is particularly relevant because it allows for more reliable estimates when examining relationships between variables (e.g., hemoglobin, weight, height) across diverse groups of children over multiple time points. This method improves the accuracy of the regression estimates by accounting for possible differences in variance across age groups and individual children, making it a robust choice given the heterogeneity in the dataset. These methods allow researchers to control unobserved heterogeneity.

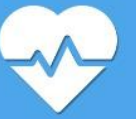

We believe that Panel Data (FGLS) methods with a relative abundance of data for the chosen variables such as hemoglobin, height, and weight, related to age segments, should provide a perspective to deduct whether the intervention plans in the fight against child malnutrition are having effects or not, even when the specific contributions of each plan may not be known. In simple terms, this method allows us to know if, in the period selected for the study, the correct "a-priori" correlations and interactions between hemoglobin, weight, and height in relation to the age of children under five years can be verified. Given the observational nature of this study, establishing causal relationships between hemoglobin levels and anthropometric measures is challenging. While the use of FGLS methods helps mitigate issues of heteroscedasticity and serial correlation, future studies could employ advanced methodologies, such as instrumental variable models or regression discontinuity designs, to better approximate causality. These methods would strengthen the evidence base for evaluating the effectiveness of public health interventions.

## 3 RESULTS AND DISCUSSION

The comparative evolution of hemoglobin, weight and height in children under five years of age from 2000 to 2022, we can see at Table 1 noticeable improvements throughout the period, although they are relatively modest and fluctuating. In the case of mean values for these variables, the hemoglobin level goes from 10.894 g/dL to 11.532 g/dL, an increase of 5.86%, while the increase in weight and height was 7% and 4.21%, respectively.

It is also interesting to see two different trends when we address standard deviation values. While it is decreasing when related to hemoglobin level measurements, which suggests that interventions to combat chronic anemia seem to be on track, these are not the cases for weight and height. Despite greater sample sizes in the ENDES survey in most of the years of the study, the greater dispersion observed in weight and height may indicate a persistence of other problems, possibly closer to deficiencies for the adequate development of children and preliminary situations of overweight and obesity.

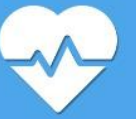

The aggregation in Table 1 is intended to enhance readability; however, the econometric analysis employs consistent age groups (0–11, 12–23, 24–35, 36–47, and 48–59 months) across all survey years. A supplementary table in the appendix confirms the relative proportionality of these groups throughout the study period.

Table 1. Composite table on average Hemoglobin (g/dL), Weight (kg) and Height (cm) as for selected years.

|  | 2000 | 2007 | 2012 | 2017 | 2022 |
|---|---|---|---|---|---|
| **Hemoglobin** | 10.894 | 10.875 | 11.309 | 11.291 | 11.532 |
| std.dev. | 1.5765 | 1.3956 | 1.2698 | 1.1431 | 1.0818 |
| n (size) | 2513 | 3648 | 8224 | 19,012 | 19,973 |
| **Weight** | 12.213 | 11.907 | 12.224 | 12.612 | 13.068 |
| std. dev. | 3.7528 | 3.6518 | 3.8341 | 3.9229 | 4.0489 |
| n (size) | 11,894 | 4777 | 9235 | 20,913 | 21,416 |
| **Height** | 83.668 | 83.067 | 84.803 | 86.022 | 87.187 |
| std. dev. | 13.611 | 13.736 | 14.024 | 13.940 | 13.753 |
| n (size) | 11,795 | 4722 | 9228 | 20,912 | 21,903 |

Source of Data: ENDES, INEI – Peru.

Using Tables 2 (from a. to e.) we can understand the evolution of anemia for selected years, based on the availability of that information by ENDES and following some of the measurements suggested by Bland (2015) and Rosner (2016). Anemia dramatically diminishes with children´s age and throughout time. For instance, in 2000 the cohort from 0 to 11 months just reached a normal level of 39.85%, while that level was 64.95% for children in the 48 to 59 months range. In 2022, the youngest group got 37.69% of normalcy, while the oldest one was measured at 76.81% for hemoglobin levels equal or higher than the cutoff (11.0 g/dL). The whole period shows clear improvements in anemia prevalence, although some critical persistence of severe and moderate anemia is still worrisome in 2022 for the two younger groups.

Table 2. a. Anemia distribution by condition and age for the year 2000

| Age (months) | Normal (%) | Mild (%) | Moderate (%) | Severe (%) | Sample size |
|---|---|---|---|---|---|
| 0–11 | 39.85 | 21.63 | 36.92 | 2.20 | 409 |
| 12–23 | 31.08 | 27.61 | 38.04 | 3.27 | 489 |
| 24–35 | 50.67 | 26.67 | 21.71 | 0.95 | 525 |
| 36–47 | 61.82 | 23.09 | 14.53 | 0.56 | 537 |
| 48–59 | 64.94 | 19.50 | 15.03 | 0.54 | 559 |

Source of Data: ENDES, INEI - Peru

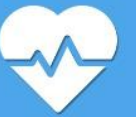

Table 2. b. Anemia distribution by condition and age for the year 2007

| Age (months) | Normal (%) | Mild (%) | Moderate (%) | Severe (%) | Sample size |
|---|---|---|---|---|---|
| 0–11 | 20.65 | 31.59 | 44.78 | 2.99 | 402 |
| 12–23 | 32.63 | 28.40 | 38.97 | 2.00 | 852 |
| 24–35 | 53.13 | 31.79 | 15.59 | 0.49 | 821 |
| 36–47 | 62.34 | 22.04 | 15.49 | 0.13 | 794 |
| 48–59 | 69.32 | 21.18 | 9.24 | 0.26 | 779 |

Source of Data: ENDES, INEI - Peru

Table 2. c. Anemia distribution by condition and age for the year 2012

| Age (months) | Normal (%) | Mild (%) | Moderate (%) | Severe (%) | Sample size |
|---|---|---|---|---|---|
| 0–11 | 32.20 | 30.11 | 36.72 | 0.94 | 847 |
| 12–23 | 44.04 | 30.08 | 24.99 | 0.89 | 1805 |
| 24–35 | 68.43 | 21.90 | 9.45 | 0.21 | 1904 |
| 36–47 | 78.30 | 17.16 | 4.49 | 0.05 | 1871 |
| 48–59 | 79.98 | 15.59 | 4.38 | 0.05 | 1828 |

Source of Data: ENDES, INEI - Peru

Table 2. d. Anemia distribution by condition and age for the year 2017

| Age (months) | Normal (%) | Mild (%) | Moderate (%) | Severe (%) | Sample size |
|---|---|---|---|---|---|
| 0–11 | 37.57 | 32.99 | 28.83 | 0.61 | 2140 |
| 12–23 | 47.87 | 31.66 | 19.95 | 0.52 | 4466 |
| 24–35 | 67.67 | 24.32 | 7.82 | 0.19 | 4182 |
| 36–47 | 73.77 | 20.93 | 5.18 | 0.12 | 4224 |
| 48–59 | 80.90 | 15.95 | 3.10 | 0.05 | 4000 |

Source of Data: ENDES, INEI - Peru

Table 2. e. Anemia distribution by condition and age for the year 2022

| Age (months) | Normal (%) | Mild (%) | Moderate (%) | Severe (%) | Sample size |
|---|---|---|---|---|---|
| 0–11 | 37.69 | 34.69 | 27.20 | 0.33 | 3051 |
| 12–23 | 50.53 | 32.00 | 17.20 | 0.27 | 4069 |
| 24–35 | 69.70 | 22.73 | 7.44 | 0.12 | 4139 |
| 36–47 | 73.11 | 22.37 | 4.52 | 0.00 | 4314 |
| 48–59 | 76.81 | 19.27 | 3.82 | 0.02 | 4400 |

Source of Data: ENDES, INEI - Peru

Consistent as said, from the graphical standpoint, Figure 1 allows us to see the full range of observations (n = 209,011) for valid hemoglobin measurements against all the groups of children for the whole period. We see how the value of hemoglobin rises through the age cohorts. From the obtained regression line (y = 10.4 + 0.0276x), based on 209,211 observations, we observe a positive relationship between age and hemoglobin levels over 2007-2022. This pattern reflects both normal physiological developments, particularly during the first year of life, and potential influences of nutritional and health improvements as children grow older. The initial decrease in hemoglobin after birth, followed by an increase during the first year, aligns with normal child development, whereas sustained increases in later years may reflect external factors such as healthcare access and dietary quality.

Figure 1. Evolution of hemoglobin levels by age (children under five 2007 – 2022

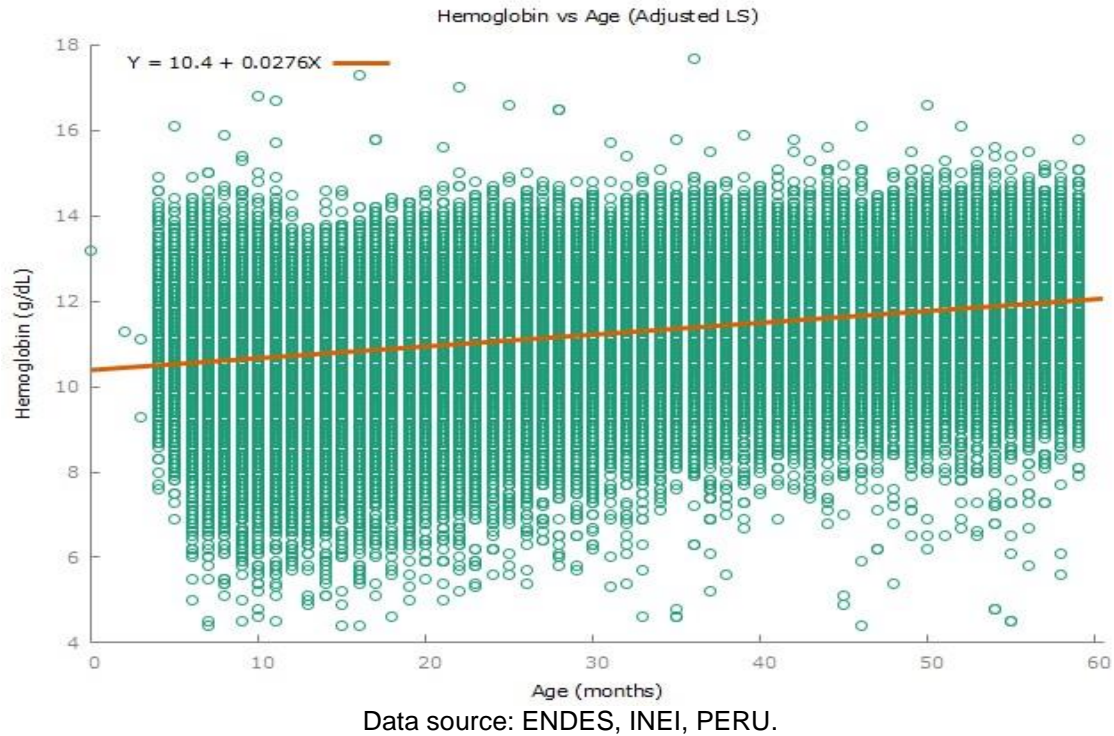

Data source: ENDES, INEI, PERU.

We may also see the improvements of height and hemoglobin together, as shown in Figure 2, where the regression line (y = 41.3 + 4.16x) provides a positive slope and is particularly useful on how height and hemoglobin share gains for the children being studied. It is dramatic, however, the large part of the figure that is below the average height-to-age in relation to the level of hemoglobin, or to the standard recommended by the WHO (2006). Further research should be undertaken for that area that comprises near 25% of the children that were evaluated by ENDES, limited to below the levels of less than 11 g/dL of hemoglobin and 80 cm in height, where stunting may prevail.

Figure 2. Evolution of height to hemoglobin for all children under five 2007 – 2022

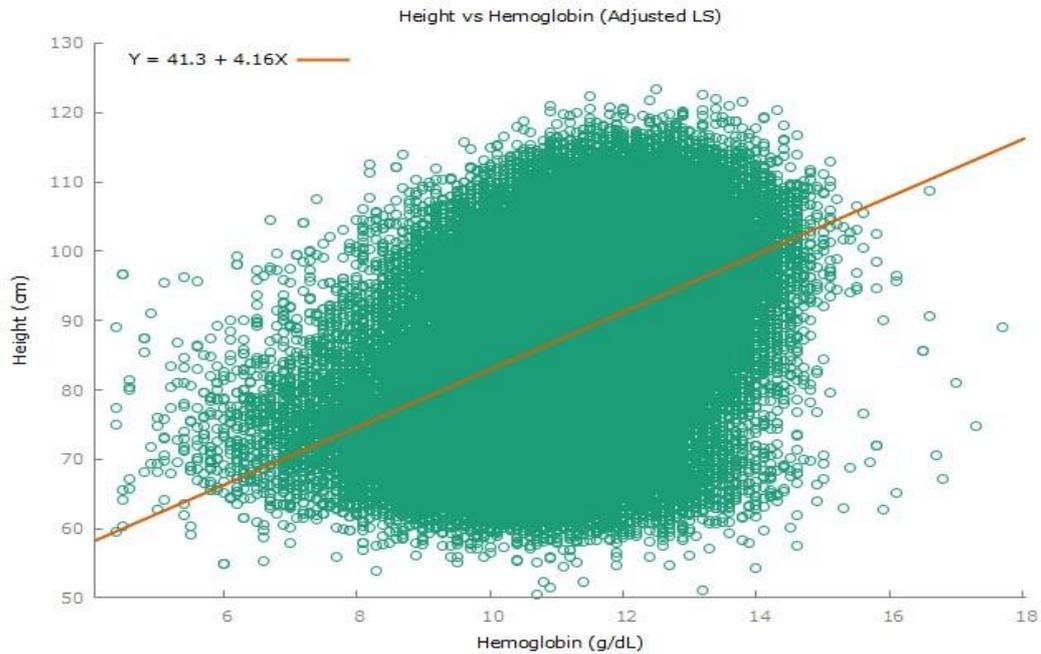

Source of Data: ENDES, INEI, Peru.

Similarly, hemoglobin and height show also a positive relationship (y = 0.023 + 1.17x) as well, although the somewhat different shape of the accumulation of pairs may lead to further research to understand different nutritional diets that may introduce certain biases regarding weight gains, especially toward child obesity.

Figure 3. Evolution of weight to hemoglobin for all children under five 2007 – 2022

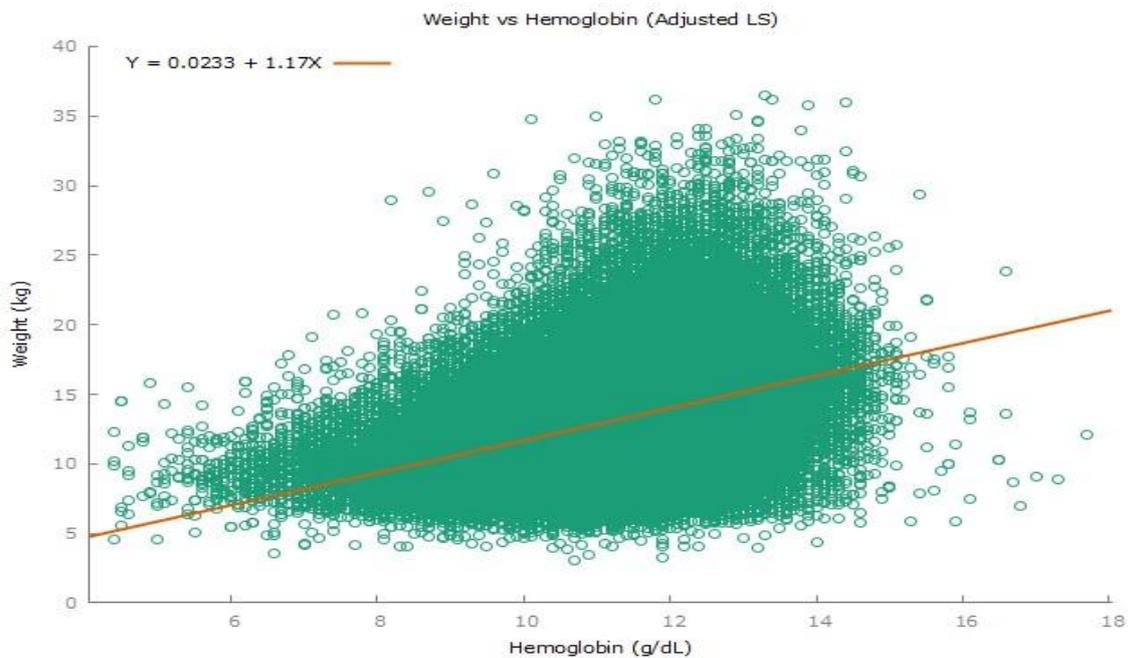

Data: ENDES, INEI, Peru.

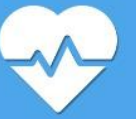

## 3.1 ECONOMETRIC RESULTS

We have employed GRETL version 2025a, for econometric estimation, and IBM´s SPSS version 29.0.0.0 (241) to process raw data from ENDES and then to convert into GRETL´s working files. After purging all the data from outliers and missing observations, the datasets under study have been 209,011 observations for the relationship between hemoglobin and age, 208,763 observations for the estimation between weight and hemoglobin, and 208,713 pairs for the one linking height and hemoglobin. Panel Data regressions are quite revealing, since they allow us to deal with a very large data set for both dimensions, namely time and cross-section. There are well-conceived papers that employ Panel Data models or statistically similar ones in this subject, as we find in Rajagukguk *et al.* (2021), or in Nigatu *et al.* (2021), with a relevant meta-analysis provided by Gedfie *et al.* (2022).

It is important to see the evolution of anemia in Peru, according to ENDES surveys. To understand the situation of normal hemoglobin levels, based on the cut-off of 11 g/dL according to WHO and the categories of anemia (mild, moderate and severe) presented in tables 2 to 6, it should be noted that throughout the period it is always the segment between 0 to 11 months where the lowest normal levels are found. Even so, it is visible how throughout the first years of life normal hemoglobin levels improve and the levels of moderate and severe anemia decrease throughout the period, although with fluctuations. These figures may also be indicative of the precarious living conditions of pregnant mothers and of the feeding of children between 0 and 11 months, while nutritional programs and health interventions of the Peruvian government have some degree of success after that age group. In our work, we provide three regression groups.

### 3.1.1 Regression group I.- hemoglobin (g/dL) vs age (months)

In this first set, we obtain robust values that represent the expected a-priori relationship between the hemoglobin and age, that is, incremental and with positive sign, except for the tranche of 0 to 11 months, which is consistent with what we have explored before. All estimates are relevant at 99% of statistical significance.

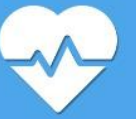

In examining the relationship between age (measured in months) and hemoglobin levels across early childhood, a series of Panel Data regressions were conducted, segmenting the sample into five distinct age groups: 0 to 11 months, 12 to 23 months, 24 to 35 months, 36 to 47 months, and 48 to 59 months. The hemoglobin level was modeled as the dependent variable, while age served as the independent variable within each age group.

The regression analysis yielded statistically significant results across all age groups, with z-statistics indicating a 99.9% confidence level in the relevance of age to hemoglobin levels, thus suggesting a robust statistical relationship. The coefficients associated with age across the regressions were -0.05432, 0.06729, 0.012879, 0.02344, and 0.17371 for each respective age group. These coefficients represent the average change in hemoglobin levels associated with a one-month increase in age within each age category.

The intercepts for these regressions, which indicate the estimated hemoglobin level at birth (0 months) for the respective age groups, were found to be 10.9853, 9.6169, 11.0504, 10.6898, and 10.8374, respectively. The intercepts in a regression model represent the estimated value of the dependent variable (hemoglobin level) when all independent variables are zero. However, in this case, "age = 0 months" does not perfectly reflect hemoglobin levels immediately after birth. The model does not account for the transiently high hemoglobin levels in neonates (16-20 g/dl), which rapidly decline during the first month due to normal physiological processes such as the transition from fetal to adult hemoglobin. These values provide a baseline from which the effect of age on hemoglobin can be understood, adjusting for the age-specific baseline differences in hemoglobin levels.

The negative coefficient observed in the youngest age group (0 to 11 months) suggests a decrease in hemoglobin levels with each additional month of age within this group. Conversely, positive coefficients in the remaining groups indicate an increase in hemoglobin levels with age. The varying direction and magnitude of these coefficients across age groups highlight the complex and dynamic nature of hemoglobin level development during early childhood.

The large sample sizes associated with each regression strengthen the reliability of these findings, providing a solid foundation for inferring the generalizability of the observed trends in hemoglobin changes with age in a broader population.

Table 3a. Regression table group I

| Age (months) | 0–11 | 12–23 | 24–35 | 36–47 | 48–59 |
|---|---|---|---|---|---|
| Constant | 10.9853 | 9.61697 | 11.0504 | 10.6898 | 10.8374 |
| Coefficient | -0.05432 | 0.067289 | 0.012879 | 0.0234406 | 0.17371 |
| Std. Error | 0.004848 | 0.001352 | 0.001649 | 0.0006438 | 0.001199 |
| z-statistic | -11.20 | 47.78 | 7.81 | 36.41 | 14.49 |
| $\rho - value$ | *** | *** | *** | *** | *** |
| Sample size | 24,169 | 45,523 | 46,686 | 46,174 | 46,459 |

Statistical significance levels are denoted as follows: *p < 0.05, **p < 0.01, ***p < 0.001.
Sources: Authors.

### 3.1.2 Regression group II.- weight (kg.) vs hemoglobin (g/dL)

This second set of observations and corresponding regressions show a robust direct relationship between these two variables, with increasing constant and coefficients, as expected.

Table 3b. Regression table group II

| Age (months) | 0–11 | 12–23 | 24–35 | 36–47 | 48–59 |
|---|---|---|---|---|---|
| Constant | 8.23515 | 7.73848 | 8.83427 | 9.75562 | 11.0391 |
| Coefficient | 0.0255723 | 0.242426 | 0.343458 | 0.448501 | 0.506997 |
| Std. Error | 0.002982 | 0.003984 | 0.013259 | 0.020501 | 0.026639 |
| z-statistic | 8.552 | 60.85 | 25.90 | 21.88 | 19.03 |
| $\rho - value$ | *** | *** | *** | *** | *** |
| Sample size | 24,162 | 45,499 | 46,094 | 46,622 | 46,386 |

Statistical significance levels are denoted as follows: *p < 0.05, **p < 0.01, ***p < 0.001.
Sources: Authors.

### 3.1.3 Regression group III.- height (cm) vs hemoglobin (g/dL)

Finally, strong estimated values are also found when the regression method is applied to height and hemoglobin. There is, however, an estimated value that does not meet the conditions of statistical relevance, and that is the coefficient corresponding to the first age group, that is, from 0 to 11 months, which has been the segment of greatest imprecision and concern. In this case, we presume that height measurements are much more difficult to perform, given that most of the children, according to the ENDES survey itself, are measured lying down and not standing, which can lead to frequent measurement errors, even though the size of the sample (24,156 observations) is quite substantial. To address causal ambiguity, lagged regression models could be employed, using previous-year hemoglobin levels as predictors of current weight and height. Preliminary results suggest stronger predictive relationships from hemoglobin levels to growth metrics, although causality cannot be definitively established. We believe that further longitudinal studies might be needed to clarify these relationships.

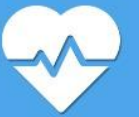

Table 3c. Regression table group III

| Age (months) | 0–11 | 12–23 | 24–35 | 36–47 | 48–59 |
|---|---|---|---|---|---|
| Constant | 68.6007 | 68.3091 | 76.5588 | 83.2327 | 90.0605 |
| Coefficient | -0.003552 | 0.898126 | 0.971841 | 1.04531 | 1.01615 |
| Std. Error | 0.011887 | 0.13115 | 0.010836 | 0.09027 | 0.036198 |
| z-statistic | -0.2976 | 68.48 | 89.69 | 11.35 | 28.07 |
| $\rho - value$ | No * | *** | *** | *** | *** |
| Sample size | 24,156 | 45,483 | 46,056 | 46,616 | 46,402 |

Statistical significance levels are denoted as follows: *$p < 0.05$, **$p < 0.01$, ***$p < 0.001$.
Sources: Authors.

## 3.2 DISCUSSION

We believe that it is possible to prove with a certain degree of statistical significance that, in the fight against malnutrition, anemia, and stunted growth in children under five years old in Peru, the various programs and interventions are achieving positive effects. However, the use of these large databases and more advanced computational methods will also allow us to make a better and more precise dissection of the samples and surveys, so that more accurate data can be recorded in those groups that are more vulnerable or where it is suspected that the actions to combat these adverse situations for children are not achieving the desired effects. There are important discussions in Gertler (2004), Hoddinott *et al.* (2009), and IMTF (2018) on how these interventions are truly mitigating or not these situations.

There are, therefore, various limitations and restrictions that we have imposed on ourselves, for now, in this study. In the current study, there is no separation by gender, which indeed limits the potential to understand gender-specific trends and outcomes in anemia, weight, and height. While this study provides valuable insights into the trends of anemia, weight, and height in children under five years old in Peru, it does not include a gender-specific analysis.

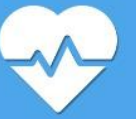

Although ENDES data does not indicate statistically significant differences in hemoglobin levels or anemia prevalence between boys and girls, future research could explore whether intervention programs equitably benefit both genders. A gender-disaggregated analysis might uncover patterns linked to biological and socio-cultural differences, offering opportunities for more tailored public health strategies. Biological factors, such as differences in growth rates and nutritional needs, as well as social factors, such as access to resources and healthcare, could vary significantly between boys and girls. Including this analysis could help policymakers tailor health programs more effectively for each gender. To continue making progress, we will be to move towards the details that allow finding findings such as those that will emerge from segmenting the samples between boys and girls, also segmenting these samples by socioeconomic level, household wealth, place of residence and other conditions (whether in an urban, peri-urban, or rural environment) as seen in Abate *et al.* (2021) for women of reproductive age in Ethiopia, as well as the educational conditions of the parents, and particularly, of the mother (see Subramanian *et al.,* 2009). For example, if it were consistently found that girls have lower levels of hemoglobin, which in turn correspond to smaller sizes and weights than those established as standards by the WHO, we should then question whether the intervention programs are keeping the basic elements of social and gender equity established by law. In the same way, if the children showing worse outcomes correspond to mothers who are in a situation of high vulnerability or who have suffered marginalization regarding access to education, to vocational training, and to means of work and a decent life, again we would be facing a problem where latent or persistent gender discrimination would be deepening the already precarious situation produced by chronic malnutrition. These types of findings have been recently provided by Mirza *et al.* (2018), Sadiq *et al.* (2018), Hlahla *et al.* (2023), and Iglesias (2019) show us how complex, persistent and extended this situation is worldwide.

However, and despite this discussion, according to ENDES reports, no statistically significant differences have been found in hemoglobin levels or in the prevalence of anemia between boys and girls aged 0 to 5 years. The available data indicates that both boys and girls have similar anemia rates, with no substantial variation that is statistically significant. For example, in the Peruvian department of

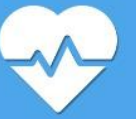

Ancash, the ENDES 2021 report shows that moderate anemia affected 8.7% of children, with a difference of 4.3 percentage points between urban (7.0%) and rural (11.3%) areas. However, these differences are attributed more to the area of residence than to the sex of the children. In summary, the available information suggests that there are no statistically significant differences in hemoglobin levels and the prevalence of anemia between boys and girls aged 0 to 5 years in Peru, according to the data collected by the ENDES.

Peru's diverse geographical and socioeconomic landscape significantly influences anemia and malnutrition prevalence. Although this study adjusts hemoglobin levels for altitude, a more detailed regional analysis could uncover disparities between urban, peri-urban, and rural areas. Future research should focus on these differences to assess how local factors, such as healthcare access and availability of nutritional supplementation programs, impact outcomes. Such stratified analyses would provide actionable insights for policymakers aiming to address region-specific challenges. While the study utilizes a large sample size from ENDES surveys, it would benefit from a discussion on the representativeness of this sample across Peru's diverse regions. But, despite these concerning facts and results mentioned above, the ENDES surveys also highlight, for instance, the deep penetration of cell phones in Peru, up to the point that practically the entire population owns or has quick access to one. In fact, the most recent figure as of June 30th, 2024, provided by the Peruvian authorities places it at 41 million active lines. The high penetration of mobile phones in Peru presents a valuable opportunity for the government to implement targeted interventions aimed at reducing anemia rates among children under five. This technological reach can facilitate the dissemination of critical nutritional information to mothers, ensuring that best practices are communicated effectively and at scale can allow for educational campaigns and outreach from public programs fighting malnutrition in children under five years old, to capture the attention that families have towards this social problem, informing them about it and receiving recommendations on diets and nutritional values, as well as on sanitary aspects of managing drinking water and waste. Strategies like preloaded messages on essential nutrition and hygiene practices can enhance the reach and impact of such campaigns. These conditions seem quite relevant, particularly for young pregnant women and their babies of less than one year of age, as we can see in Maulide *et al.* (2022) and Scott *et al.* (2014), since we observed negative or near zero coefficients from the Panel Data regressions in the cohort of 0 to 11 months. In future research, we will address the points of discussion explained above. To enhance the robustness of the model, incorporating control variables could help isolate the effect of the main variables (anemia, weight, height) on child health outcomes. Potential control variables might include, a) Access to healthcare services: Distance to medical centers or availability of healthcare workers could impact intervention effectiveness; b) Parental education levels: Higher education levels often correlate with better child health outcomes; c) Socio-economic status (SES): Income level, type of housing, or ownership of durable goods can act as proxies for SES, which can influence nutrition and healthcare access. Physiotherapy may also play a crucial role in the rehabilitation of children with malnutrition and growth retardation, especially through early stimulation programs. Studies such as those developed by Humanity & Inclusion have shown that stimulation therapy improves the psychomotor and cognitive development of these children, complementing nutritional recovery. The integration of these interventions in public health programs can optimize long-term outcomes, reducing functional sequelae and promoting a better quality of life.

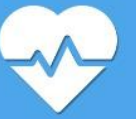

There is evidence supporting the efficacy of physiotherapy in the rehabilitation of malnourished children. A study conducted by Humanity and Inclusion in Mali during year 2014 (see 2021) evaluated the short-term effects of stimulation therapy in the treatment of severe acute malnutrition in children aged 6 months to 5 years. The results showed significant improvements in the psychomotor and cognitive development of children who participated in the stimulation program compared to those who did not receive such an intervention.

In addition, a comparative study of nutritional recovery in Bolivia by Sevilla *et al.* (2010) analyzed 176 malnourished children, of whom 146 participated in a rehabilitation program that included stimulation and physiotherapy components. The children who participated in the program showed a faster and more complete recovery compared to those who did not participate in the program. Including these control variables in our future regression models would help account for confounding factors, enhancing the clarity of the relationships we are examining.

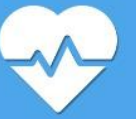

In the case of this study, it analyzes trends in hemoglobin, weight, and height over time, which may reflect the cumulative impact of various public health policies. While trends in hemoglobin, weight, and height may correlate with public health interventions, this study does not establish direct causal links. Observations remain suggestive and require further experimental or longitudinal data to substantiate claims regarding the specific impacts of government programs. However, quantifying the specific contributions of individual government programs requires more targeted evaluation models, such as difference-in-differences regressions or cohort analyses. Future studies should aim to disentangle the differential impacts of these interventions to better understand their effectiveness and optimize resource allocation.

## 4 CONCLUSIONS

Anemia continues to be a major health problem, especially among women and children under five years of age in less developed countries, with Latin America being one of the regions of greatest interest for interventions. This study examines the relationship between age, hemoglobin levels, weight, and height in children under five in Peru from 2007 to 2022, leveraging a large dataset and advanced econometric methods. While the analysis reveals positive trends in these parameters over the years, it is essential to interpret these findings within the biological and physiological context of child development. The observed increase in hemoglobin levels with age aligns with established physiological trends, where hemoglobin naturally declines after birth and gradually increases as the child's hematopoietic system matures. While improved healthcare access and nutritional interventions may contribute to sustaining this increase in later years, the early life patterns primarily reflect normal biological development. The positive correlation between age and anthropometric indicators (weight and height) underscores the influence of growth patterns. These trends suggest potential improvements in nutritional and health conditions over time but also reflect the inherent growth trajectory of young children.

After more than 20 years of fighting against anemia and chronic malnutrition among children under 5 years old, Peru can say that it is on the right path towards reducing this situation. The prevalence of anemia has decreased. However, the persistence of certain levels of anemia, malnutrition, and delays in the proper growth of the child population are still concerning. It may be advisable to extend these concepts and those used by the WHO regarding anthropometric measures to the entire affected population. As mentioned above, the high penetration and presence of the cell phone in almost the entire Peruvian population should be a tool for the dissemination of ideas and concepts for better and more nutritious feeding, within the economic possibilities and procurement of households. The focus should be on the first stage of life, that is, between 0 and 11 months, where the expected results have not yet been achieved. On that last statement, this study is limited by the lack of physiological data and direct measures of policy effectiveness. Therefore, collaborating with pediatricians and nutrition experts in future research could provide a more nuanced understanding of the interplay between biological development and external interventions. Additionally, incorporating randomized trials could help establish causality and refine policy recommendations.

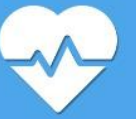

There are two limitations in this research that we need to share with the readers. The first one is the lack of separation between boys and girls with respect to the statistical and econometric procedures adopted. The main reason has been a preliminary finding of no statistical difference, from the ENDES data. We think, however, that we need to go deeper into regional and social data to see if that aggregate information of no gender difference persists. The other limitation to replicate this study in developing countries could be the volume and the quality of data. We have been quite fortunate to what we found in ENDES (INEI, Peru), but that could not be the case in other countries. Our desire is to extend this research to more developing nations, as well as we recommend scholars to introduce more socioeconomic and environmental variables and to additionally apply methods of cause-effect for measurement and evaluation of policy interventions.

## ACKNOWLEDGEMENTS

We are grateful for the valuable comments of colleagues in our institutions. Additionally, we would like to thank Dr. José-Miguel Martínez-Carrión of the

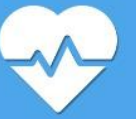

University of Murcia, Spain and the project ref. PID 2020-113793GB-I00 (MCIN/AEI/10.13039/501100011033 - Government of Spain). Finally, we thank the Fundación Universidad Alfonso X El Sabio (Madrid, Spain) and the Center for International Policy Studies – CIPS, of Fordham University, New York, USA for their continuous support.

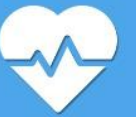

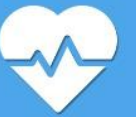

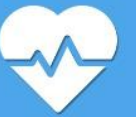